# INJECTION AND DUMP CONSIDERATIONS FOR A 16.5 TEV HE-LHC

B. Goddard, M. Barnes, W. Bartmann, J. Borburgh, C. Bracco, L. Ducimetière, V. Kain, M. Meddahi, V. Mertens, V. Senaj, J. Uythoven, CERN, Geneva, Switzerland

*Abstract*

Injection and beam dumping is considered for a 16.5 TeV hadron accelerator in the current LHC tunnel, with an injection energy in the range 1 – 1.3 TeV. The present systems are described and the possible upgrade scenarios investigated for higher beam rigidity. In addition to the required equipment performance, the machine protection related aspects are explored. The expected constraints on the machine layout are also given. The technological challenges for the different equipment subsystems are detailed, and areas where R&D is necessary are highlighted.

## ASSUMED PARAMETERS

The most important parameters assumed for injection and extraction from HE-LHC are listed in Table 1, with the values for LHC (ultimate bunch intensity) also included. The resulting total energy per transfer (injection or extraction) and beam sizes are also included.

Table 1. Assumptions of target parameters for HE-LHC.

|  |  | LHC | | HE-LHC | |
| --- | --- | --- | --- | --- | --- |
|  |  | Inject | Extract | Inject | Extract |
| Bunch I | p+ | 1.7e11 | 1.7e11 | 1.3e11 | 1.3e11 |
| p·c | TeV | 0.45 | 7 | 1.3 | 16.5 |
| Rigidity | Tm | 1503 | 23337 | 4337 | 55004 |
| Nb/transfer |  | 288 | 2808 | 144 | 1404 |
| E/transfer | MJ | 3.5 | 535 | 3.9 | 482 |
| $\varepsilon_{xyn}$ | µm | 3 | 3 | 2 | 2 |
| $\beta_{x/y}$ (septum) | m | 100 | 100 | 100 | 100 |
| $\beta_{xy}$ (dump) | m | 4500 | 4500 | 4500 | 4500 |
| $\sigma_{xy}$ (septum) | mm | 0.79 | 0.20 | 0.38 | 0.11 |
| $\sigma_{xy}$ (dump) | mm | 5.30 | 1.34 | 2.55 | 0.72 |
| Injection gap | µs | 1.00 | 1.00 | 1.00 | 1.00 |
| Dump gap | µs | 3.00 | 3.00 | 3.00 | 3.00 |

## INJECTION AT 1.3 TEV

### Present injection system

The present fast injection systems in P2 (Beam 1) and P8 (Beam 2) use DC powered laminated steel Lambertson septum magnets and ferrite yoke transmission line pulsed extraction kicker magnets. The kickers use thyratron switches and have ceramic vacuum chambers which support screening elements to reduce the beam coupling impedance.

The injection systems also comprise beam instrumentation and dedicated passive protection devices to intercept beam in case of an injection kicker failure. Overviews and details of the systems and components can be found in [1,2].

The injection systems for Beam 1 and Beam 2 are located to the left of P2 and right of P8, respectively, and are integrated into the matching sections of the low-β insertions for the ALICE and LHCb experiments. This cohabitation imposes some optics constraints and also has proven to introduce operational complications with the background and beam losses at injection.

### Assumptions

It is assumed that the existing transfer line tunnels from the SPS will be used, with superconducting magnets for the 1.3 TeV beam transfer, and also that the insertions in P2 and P8 will continue to house the physics experiments and low-β insertions, although the insertion layout and optics may change.

### Injection kicker considerations

A major issue with 1.3 TeV injection is the strength of the injection kicker. The system will require a similar deflection to the present 0.8 mrad (this maybe be reduced by 10-15% by changes in the optics and increasing the kicker-septum drift, but probably no more). The present system is already very pushed in terms of performance, and is considered to be at the technological limit with the 60 kV switches, cables, pulsed HV insulation in vacuum and beam screens. There is no extra space in the present layout, and clearly any solution which modifies the layout is complicated by the requirement to combine the injections and experiments in the same insertions. The possible options are considered. In both cases it is assumed that the horizontal magnet gap can be reduced from 52 to 42 mm.

The first option is to increase the installed kicker length from 16.9 m to around 34 – 40 m. This would imply 40 – 46 m spacing between adjacent quadrupoles, compared to the present 22 m, and a completely new insertion layout and optics.

The second option is to double the present 1 µs rise time to around 2 µs, e.g. with the kickers in short circuit mode or with lower impedance. The installed length could then be kept to around the present length of 17 m. The ferrite saturation might be an issue with a peak field in the ferrite of about 0.26 T. This would reduce the number of bunches per injection by about 20, or about 10% of the total number of bunches. To partially compensate for this reduction, it would be rather simple to increase the Pulse Forming Network (PFN) and kick pulse length from 8 to up to 16 µs (also for the SPS extraction kickers), although this might pose other limitations in the SPS with a much higher total intensity.

The system parameters for these two options are compared with the present system in Table 2.

Table 2. Parameters for 1.3 TeV injection kicker system options. Critical values are shown in red.

| Parameter | Unit | LHC nominal | More magnets | Longer rise time |
|---|---|---|---|---|
| H gap | mm | 54 | 42 | 42 |
| Rise time | µs | 1.00 | 1.00 | **2.20** |
| Angle | mrad | 0.8 | 0.8 | 0.8 |
| ∫B.dl | Tm | 1.2 | 3.5 | 3.5 |
| Gap field | T | 0.08 | 0.11 | 0.24 |
| Peak field | T | 0.09 | 0.12 | **0.26** |
| dI/dT | kA/µs | 5.40 | 5.39 | 5.51 |
| Imax | kA | 5.4 | 12.1 | 12.1 |
| Lmag | m | 14.6 | **32.9** | 14.6 |
| Filling | | 0.864 | 0.864 | 0.864 |
| Ltotal | m | 16.9 | **38.0** | 16.9 |
| #magnets | | 4 | **9** | 4 |

Other aspects which would need detailed consideration for a new design would include impedance, beam screens, magnet core heating and electron cloud.

### Injection septum considerations

The injection septum design is very similar to the dump septum (see later). If stronger units cannot be designed then more magnets and hence more space will be needed. The present 22 m installed length would need to increase to about 43 – 55 m. With the 22 m drift needed to clear the upstream cryostat, this imposes 50 – 60 m drift between the quadrupoles surrounding the septum.

### Injection protection considerations

As for the dump protection, the injection protection device design increases in difficulty for 1.3 TeV. The protection in the transfer lines and the protection against the kicker failures would need redesign, for the 4 MJ and 1.3 TeV energy. The kicker protection devices would increase in length from 4 to maybe 6 – 8 m.

## DUMPING THE 16.5 TEV BEAM

### Present LHC beam dump

The present LHC beam dump uses a sequence of extract → dilute → absorb to abort the ~500 MJ beam, in a 'loss-free' way. The system comprises laminated steel pulsed extraction kicker magnets, DC powered laminated steel Lambertson septum magnets, laminated steel pulsed dilution kicker magnets and a 7.7 m long, 0.7 m Ø C cylinder forming the beam dump block, surrounded by steel and concrete shielding. Both extraction and dilution kickers use the same solid state Fast High Current Thyristor switch technology. The dump kickers have ceramic vacuum chambers with a few µm of Ti coating for reducing the beam coupling impedance.

Beam instrumentation and dedicated passive protection devices to intercept beam in case of a kicker error complete the dump system. More details on the system and components can be found in [3,4].

The total length of the beamline from extraction kicker to dump block is about 975 m. The dump block is separated from the vacuum of the beamline and the LHC by a 15 mm thick carbon composite (CC) entrance window, which for vacuum tightness has a 0.2 mm thick steel backing foil. The dump systems for Beam 1 and Beam 2 are located symmetrically about P6 of the LHC, and use the full straight section, with a special optics to provide the long drift distance needed between kicker and septum, and from the septum to the next machine quadrupole to allow the beam to be extracted past the cryostat. For the layout, there are only two stand alone matching quadrupoles each side of the IP (Q4 and Q5) which are not in the continuous cryostat. A schematic layout of the elements in P6 is shown in Figure 1.

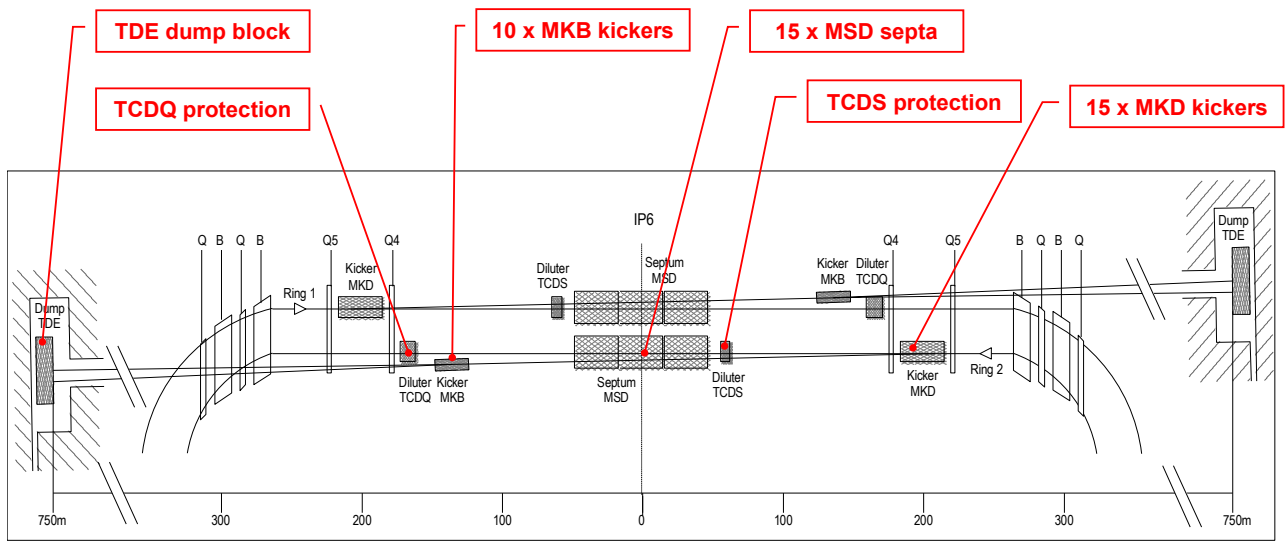

Figure 1. Schematic layout of dump elements in P6.

*Assumptions*

It is assumed that the existing tunnel and caverns are reused, Figure 2, which implies the same (similar) extraction trajectories in horizontal and vertical, and therefore similar kicker and septum angles. A maximum ~300 mm dilution sweep radius is permitted, otherwise extra civil engineering would be needed to enlarge the TJ62 and TJ68 junction caverns with the LHC. The insertion layout and optics could change if required.

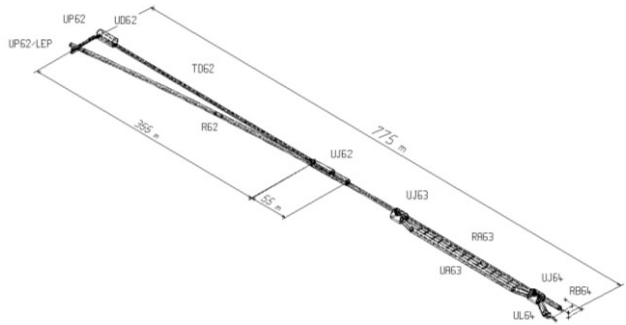

Figure 2. Underground layout for the left side of P6, showing the dump tunnel and UD62 cavern for Beam 2.

*Extraction kicker with 3 µs rise time*

An option for the extraction kicker system would be to keep 3 µs rise time and similar magnets. The required ∫B.dl could then be obtained by increasing the number of magnets, or increasing the current per magnet. These are compared to the present situation in Table 3.

A total of 35 magnets would be required if the parameters per magnet are unchanged – this would imply an installed length of 64.4 m, which is not compatible with retaining a similar optics; the distance between Q4 and Q5 would have to be much larger.

Table 3. Parameters for 3 µs rise time beam dump extraction kicker system. Critical values are shown in red.

| Parameter | Unit | LHC Nominal | More magnets | Higher current |
|---|---|---|---|---|
| V gap | mm | 72 | 72 | 72 |
| Rise time | µs | 3.00 | 3.00 | 3.00 |
| Angle | mrad | 0.27 | 0.27 | 0.27 |
| ∫B.dl | Tm | 6.3 | 14.9 | 14.9 |
| Gap field | T | 0.30 | 0.30 | 0.71 |
| Peak field | T | 0.41 | 0.41 | 0.95 |
| dI/dT | kA/µs | 6.17 | 6.23 | **14.53** |
| Imax | kA | 18.5 | 18.7 | **43.6** |
| Vmax | kV | 30.0 | 30.0 | **70.7** |
| Lmag | m | 21.0 | **49.0** | 21.0 |
| Filling | | 0.761 | 0.761 | 0.761 |
| Ltotal | m | 27.6 | **64.4** | 27.6 |
| # magnets | | 15 | **35** | 15 |

For higher current per magnet with a similar magnetic length to present, the peak current would increase to 43.6 kA and the peak field to 0.95 T, which may be about feasible. However, the dI/dt increases to 14.5 kA/µs, which requires that the system voltage increases enormously, from 30 kV to over 70 kV. This is simply not possible with the air insulated generators and switches which form the core of the system. A change to an oil insulated system would introduce many complications with footprint, maintenance and complexity and also safety issues with large quantities of oil underground. The maximum oil-insulated system voltage of around 60 kV would still not be enough to fit the system in a similar length to the present system. In addition a current feedthrough for 44 kA would be very challenging.

*Extraction kicker with longer rise time*

The second option for the kicker system would be to use longer rise time and a different magnet design. The vertical gap could be reduced to take advantage of the smaller beam size at 1-1.3 TeV injection energy. In a first assumption it is assumed that the clear vertical vacuum chamber aperture can be reduced from 62 to 42 mm, which is a factor √(450/1000). This may be slightly optimistic for 1 TeV as the allocation for orbit, mechanical and alignment tolerances are fixed quantities. The ceramic chamber and associated tolerances require another 10 mm opening, so the final vertical gap between the poles would be reduced from the present 72 mm to 52 mm. Limiting the dI/dt to the present value (which may be slightly pessimistic), the parameters obtained are shown in Table 4.

Table 4. Parameters for beam dump extraction kicker system with smaller gap and longer rise time. Critical values are shown in red.

| Parameter | | LHC Nominal | Smaller gap |
|---|---|---|---|
| V gap | mm | 72 | **52** |
| Rise time | µs | 3.00 | **5.10** |
| Angle | mrad | 0.27 | 0.27 |
| ∫B.dl | Tm | 6.3 | 14.9 |
| Gap field | T | 0.30 | 0.71 |
| Peak field | T | 0.41 | 0.95 |
| dI/dT | kA/µs | 6.17 | 6.17 |
| Imax | kA | 18.5 | **31.5** |
| Vmax | kV | 30.0 | 30.0 |
| Lmag | m | 21.0 | 21.0 |
| Filling | | 0.761 | 0.761 |
| Ltotal | m | 27.6 | 27.6 |
| # magnets | | 15 | 15 |

In this configuration the extraction requires a 5.1 µs abort gap which reduces the number of bunches in the machine by about 42 at 50 ns spacing, or about 3%. The maximum current would be almost 32 kA, which would require R&D on high current switches and high current feedthroughs, but should be feasible. This system would still be air insulated and would operate at 30 kV.

*Beam dump block considerations*

For the beam dump block, a full study would be needed to analyse the extra dilution required from the MKB kicker system for a 16.5 TeV beam. In the absence of such a study, some simple scaling considerations can be made.

The peak p+ density will be a factor ~2.4 times higher, for similar beta functions at the dump block. The shower maximum will be further into the dump block, and the transverse shower extent at the shower maximum is assumed to be independent of the transverse beam size, which may be slightly optimistic. The total energy impacting the dump is about 500 MJ, similar to the LHC ultimate beam. The sweep length should therefore be similar to the present value of 100 cm. The present block is 7.7 m long and composed of carbon with different densities of 1.73 and 1.1 g cm$^{-3}$. A longer block with lower densities is likely to be required, or at least a different grading of the existing carbon densities. The longitudinal space exists in the present 25 m UD caverns.

*Dilution kicker upgrade options*

Assuming a 100 cm long sweep length at 16.5 TeV would require 2.3 times the present ∫B.dl. However, the iron of the magnet cores of the MKB dilution kickers are already near saturation, with 1.52 T peak, so it is not possible to increase the field per magnet. The apertures are determined (to first order) by the required sweep length and failure cases, and not the beam size, such that it is not possible to greatly reduce the magnet gaps. The magnets are already under vacuum with no chamber, which means nothing can be gained here in the gap size (some small optimisation could be possible with 2 families per plane with different openings).

Again two options, Table 5, are possible to increase the dilution kicker ∫B.dl – the first is to increase the number of installed magnets, keeping the switch voltage at 30 kV. 22 magnets would be needed compared to the present 10, requiring the installed space to increase from 23 to 50 m. This might be possible from an integration point of view, as the machine is not very crowded in this vicinity. The present 10 magnets are installed on the extracted beam line in the long drift space between the extraction septa and Q4, Figure 3. Space for generators in the adjacent galleries might be more problematic.

The second option would be to increase the frequency of the sweep from 14 to 28 kHz, reducing the strength and using 18 magnets in total. The system voltage could be kept at 30 kV and a sweep length of 100 cm achieved. The total installed length would then be 41 m. Damping of the diluter kicker currents is needed to achieve a spiral, which means the sweep will cross at one point on the dump block; possible waveforms and resulting sweep are shown in Figures 4 and 5. The temperature profile and dynamic mechanical stresses in the dump block would need to be evaluated.

As the magnet core is not saturated, this second solution would also have the advantage that developments in switch and insulation technology could allow an increase of the switch voltage beyond 30 kV, with a concomitant reduction in the number of magnets required. For instance, being able to increase the system voltage to 40 kV would result in a peak field of 1.3 T and a reduction to about 14 magnets.

Table 5. Parameters for dump dilution kicker system for more magnets of the present type, and for a system with higher frequency.

| Parameter | Unit | LHC | More magnets | Higher frequency |
| --- | --- | --- | --- | --- |
| F | kHz | 14.0 | 14.0 | 28.0 |
| Angle | mrad | 0.27 | 0.27 | 0.135 |
| ∫B.dl | Tm | 6.3 | 14.9 | 7.4 |
| Field | T | 1.13 | 1.21 | 0.74 |
| Peak field | T | 1.52 | 1.63 | 0.99 |
| Voltage | kV | 22.30 | 23.89 | 29.20 |
| Current | kA | 25.0 | 26.8 | 16.4 |
| Lmag H+V | m | 11.2 | **24.6** | **20.2** |
| Filling | | 0.49 | 0.49 | 0.49 |
| Ltotal | m | 22.9 | **50.3** | 41.1 |
| #magnets | | 10 | 22 | **18** |

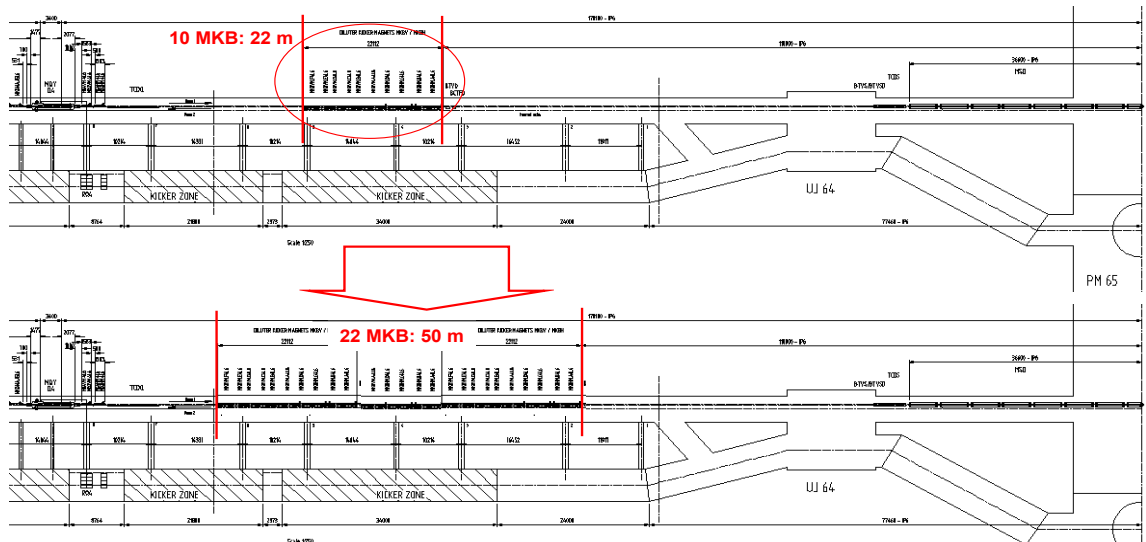

Figure 3. Present layout R6 with 10 dilution kickers (uper) and with 22 kickers (lower).

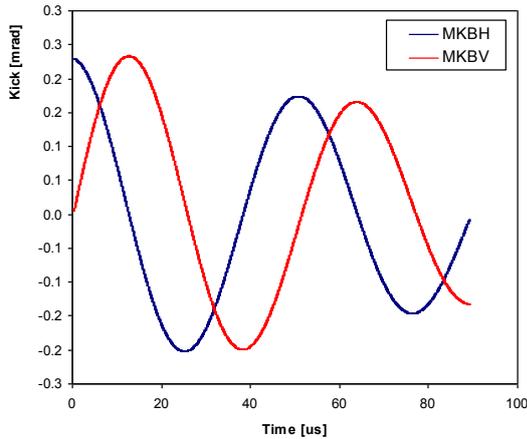

Figure 4. Possible waveforms for higher frequency sweep.

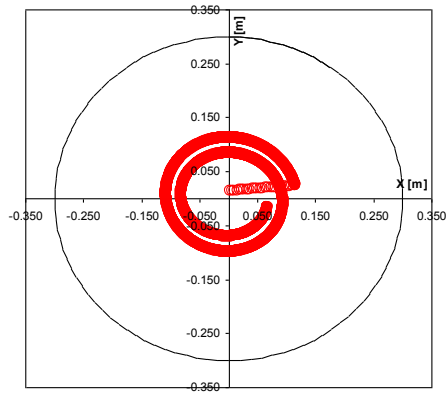

Figure 5. Sweep form on the dump block with 100 cm length and 28 kHz frequency. The sweep crosses itself, which will give locally a higher temperature in the dump.

*Dilution with a quadrupole in the beamline*

An option to supplement the dilution kickers would be to install a quadrupole or a quadrupole doublet in the beamline. If located upstream of the dilution kickers this could provide a larger beta function at the dump block. If downstream a doublet might also give a kick enhancement in both planes, increasing the effective kick strength.

The present beta values are 4-5 km at the dump. Betas of 12 km would increase the sigmas at 16.5 TeV to the present values, although this might not help the peak energy deposition if the transverse size at the shower maximum does not depend strongly on the beam size. To obtain this beta would require a 6 m quadrupole with gradient of about 150 T/m and 100 mm full aperture. The resulting line optics are shown in Figure 6.

One issue could be the trajectory offsets introduced from LHC orbit changes – with this arrangement a 4 mm orbit offset would give an additional kick of 45 μrad, producing 30 mm offset at the dump, assumed to be 650 m from the quadrupole. This should be possible to accommodate in the present 600 mm diameter dump line.

Integration of such a quadrupole is likely to be difficult upstream of the dilution kicker magnets.

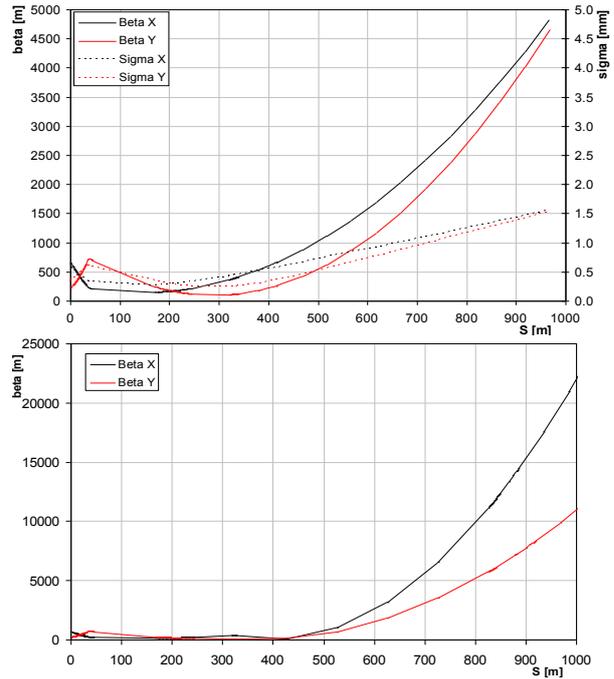

Figure 6. Present dump line optics (upper) and possible optics with 6 m long, 150 T/m dilution quadrupole (lower).

*Extraction septum considerations*

Extraction of the beam is made vertically, above the continuous cryostat. The present design uses Lambertson septa with three different septum thicknesses. For an upgrade only types B and type C would be used, as the thinnest septum is not needed behind the dedicated protection device. The field could also possibly be increased to the maximum possible value. The total number of magnets needed would then increase from 15 to 28, Table 6, and the total installed length from 73 m to 136 m. This would be difficult (although maybe not impossible) to integrate in the layout, as the drift between extraction kicker and septum entrance would be reduced by 30 m, which could in turn mean that more kick strength is required.

Alternatives are limited. Beamloss at extraction is inevitable, and so it may not be possible to build a superconducting septum. A superferric septum seems superficially interesting, to reach fields of around 2 T; however, as with all septa, saturation of the iron in the septum will strongly affect the field quality for the circulating beam, and operation above the present peak of 1.2 T may not be feasible. More studies would be needed on the septum to investigate possible alternative concepts, including ideas such as a massless superconducting septum [5].

Table 6. Extension of present extraction septum system for 16.5 TeV by increasing number of magnets.

|  | Unit | LHC | HE-LHC |
|---|---|---|---|
| Angle | mrad | 2.4 | 2.4 |
| ∫B.dl | Tm | 56.0 | 132.0 |
| Nominal field | T | 0.84 | 1.06 |
| Lmag | m | 66.7 | **124.8** |
| Filling factor |  | 0.916 | 0.916 |
| Ltotal | m | 72.8 | **136.2** |
| #magnets |  | 15 | **28** |

*Dump protection devices*

Failure cases of the extraction system include asynchronous beam dumps and high beam population in the abort gap. Protection of the septum, of the first machine quadrupole Q4 and of the collimation system from the 7 TeV beam imposes the use of long (6 m) low density (C) absorbers to intercept undiluted bunches.

The absorbers have to be low density to avoid material damage; for 16.5 TeV the densities will need to be reduced and the total length of material increased to dilute the energy density. Very long objects will be needed. A dilution factor of about $10^7$ is needed, which imposes 16 $\lambda r$ of C, which gives about 6 m of 1.8 g cm$^{-3}$ at 7 TeV. For 16.5 TeV, with smaller spot size and more energy deposited a density as low as 0.6 – 0.8 g cm$^{-3}$ may be needed to avoid damage, which would give an absorber 14-16 m long. For the fixed absorber in front of the septum this would reduce the aperture available and increase slightly the kick angle needed; for the mobile absorber in front of Q4 the challenge would be mechanical, as the absorber jaw needs to move in as the energy is ramped. Some optimisation with graded density may be possible to get more $\lambda r$ to reduce the length somewhat.

If the asynchronous dump events remain very rare (fewer than one per year, for example), an alternative would be to build sacrificial absorbers which would be damaged by a full intensity dump, and which would be easily replaceable in the event of damage. Such an option would allow shorter devices to be built, but would require R&D into failure modes and tests in a facility such as HiRadMat to check the calculations and prototypes.

## OTHER KICKER SYSTEMS

The tune kickers are not a concern for 16.5 TeV operation. They are weak devices without strong constraints on rise time, and are presently single magnets with multiple functions (several generators). It would be simple to add more kicker modules and to separate the functions, and there are no serious space constraints.

## CONCLUSIONS

Injection into HE-LHC at 1.3 TeV will need a completely new injection region layout. A longer kicker rise time of 2.2 μs with a longer pulse is feasible, as is a 1 μs rise time with 8 – 9 magnets and 40 m spacing between the adjacent quadrupoles. The number of septa would need to be doubled if stronger units could not be developed (which seems difficult), and this would require 60 m spacing between the quadrupoles. The injection protection devices also need more space. New layout and optics designs need to be investigated, which clearly have to respect the constraints coming from the experiments in injection regions.

A dump system for 16.5 TeV looks to be feasible in a similar layout to the present system. A 5 μs kicker rise time is possible with the present total system length, whereas maintaining 3 μs would require a doubling of the installed length and a major change to the insertion layout and optics. Increasing the extraction septum ∫B.dl requires an increase in septum length by a factor '1.9. This seems possible if the integration issues can be addressed. The best way of increasing the dilution sweep length seems to be by doubling the system frequency to 28 kHz. More dilution or the addition of a superconducting dilution quadrupole (or doublet) also could prove viable alternatives. The upgrade of the dump block would be rather straightforward, whereas changes to the protection devices would need much more study and development.

*Potential or required areas of R&D*

These first considerations of injection and dump systems for HE-LHC give an idea of the possible R&D directions which would be required, or which could significantly reduce other constraints on layout, optics etc.
Areas which need studies, simulations or equipment R&D (and of course the accompanying resources) are:
- Injection layout and optics;
- Dump layout and optics;
- Kicker beam screens and impedance;
- HV insulation under vacuum above 60 kV;
- Low inductance HV cables above 60 kV;
- High saturation (0.3 T) low-loss ferrites;
- High current (>40 kA) pulsed feedthroughs;
- High voltage, high current, fast solid state switches;
- Higher frequency damped generator design;
- Protection devices (low density, high strength, sacrificial designs);
- Dilution with SC quadrupoles and kickers;
- High field, beam loss resistant septa (possibly with SC or SF design).